\documentclass[aip, amsmath,amssymb, reprint, superscriptaddress]{revtex4-1}

\usepackage{graphicx}
\usepackage{dcolumn}
\usepackage{bm}

\usepackage[utf8]{inputenc}
\usepackage[T1]{fontenc}
\usepackage{mathptmx}
\usepackage{etoolbox}
\usepackage{hyperref}

\preprint{AIP/123-QED}

\makeatletter
\def\@email#1#2{%
  \endgroup
  \patchcmd{\titleblock@produce}
           {\frontmatter@RRAPformat}
           {\frontmatter@RRAPformat{\produce@RRAP{*#1\href{mailto:#2}{#2}}}\frontmatter@RRAPformat}
           {}{}
}%
\makeatother

\begin{document}

\preprint{AIP/123-QED}

\title[Magnetic Diode Effect in Skyrmion Systems]{Magnus Induced Magnetic Diode Effect in Skyrmion Systems}

\author{J. C. Bellizotti Souza$^*$}
\email{jcbsouza@dac.unicamp.br}
\affiliation{``Gleb Wataghin'' Institute of Physics, University of Campinas, 13083-859 Campinas, S\~ao Paulo, Brazil}

\author{C. J. O. Reichhardt}
\affiliation{
  Theoretical Division and Center for Nonlinear Studies,
  Los Alamos National Laboratory, Los Alamos, New Mexico 87545, USA
}

\author{C. Reichhardt}
\affiliation{
  Theoretical Division and Center for Nonlinear Studies,
  Los Alamos National Laboratory, Los Alamos, New Mexico 87545, USA
}

\author{A. Saxena}
\affiliation{
  Theoretical Division and Center for Nonlinear Studies,
  Los Alamos National Laboratory, Los Alamos, New Mexico 87545, USA
}

\date{\today}

\begin{abstract}
  We show that skyrmions can exhibit what we call a magnetic diode effect, where there is a nonreciprocal response in the transport when the magnetic field is reversed. This effect can be achieved for skyrmions moving in a channel with a sawtooth potential on one side and a reversed sawtooth potential on the other side. We consider the cases of both spin-transfer torque (STT) and spin-orbit torque (SOT) driving.
When the magnetic field is held fixed,
the velocity response of the skyrmion is the same for current applied in either direction for both STT and SOT driving, so there is no current diode effect.
When the magnetic field is reversed, under
STT driving the velocity of the skyrmion reverses and its
absolute value changes.
Under SOT driving, the velocity remains in the same direction but drops
to a much lower value, resulting in
negative differential conductivity.
For a fixed current, we find a nonreciprocal skyrmion velocity as a function of positive and negative applied fields, in analogy to the velocity-current curves observed in the usual diode effect. The nonreciprocity is generated by the Magnus force, which causes skyrmions to interact preferentially with one side of the channel. Since the channel sides have opposite asymmetry, a positive magnetic field can cause the skyrmion to interact with the ``hard'' asymmetry side of the channel, while a negative magnetic field causes the skyrmion to interact with the easy asymmetry side. This geometry could be used to create new kinds of magnetic-field-induced diode effects that can be harnessed in new types of skyrmion-based devices.
\end{abstract}

\maketitle

\section{Introduction}

In a diode effect, the transport is asymmetric when driving is applied in opposite directions. The best known example is the nonreciprocal current-voltage curves in solid-state diodes, which are used as fundamental building blocks for a wide
range of devices \cite{Shockley49}.
Diode and rectification effects are very general and can arise in other systems with some form of asymmetric geometries, such as fluids in asymmetric channels \cite{Groisman04}
and for particles driven over an asymmetric substrate where the barrier to motion in one direction is higher than the other. Diode effects have also been studied for vortices in type-II superconductors \cite{Yu07,Lu07,Lyu21},
colloids in confined geometries \cite{Reichhardt18},
and discrete charged systems \cite{Zakharov19}.
Skyrmions are particle-like magnetic textures that can be stabilized in magnetic fields and driven or manipulated by various methods, such as an applied current \cite{Nagaosa13,EverschorSitte18}.
Their motion can also be guided by manufactured defect sites or barriers \cite{Reichhardt22}.
Due to their properties, skyrmions have been attracting growing attention for numerous applications,
including novel logic devices  \cite{Zhang15,Song20,Li21},
and there have been works exploring geometries where skyrmion diode and rectification effects can occur when there is some kind of symmetry breaking \cite{Jung21,Shu22,Xu23,Souza25},
asymmetric channels \cite{Souza21,Souza22,Souza22a},
or for skyrmions moving on two-dimensional antisymmetric substrates \cite{Reichhardt15}. Similar
ratchet effects have also been studied in skyrmionium \cite{Wang20,Souza25} and hopfions \cite{Souza25a}.
Typically, in particle-like systems, if the overall substrate is symmetric, diode and rectification effects will not occur.
Additionally, in most systems with asymmetric substrates, if the direction of the magnetic field is reversed, the diode effect is unmodified. An open question is whether there can be a magnetic analog of a diode effect where a nonreciprocal response occurs for opposite magnetic field directions rather than opposite driving directions.

A prominent feature of skyrmions is that they have a Magnus component to their dynamics. Under an applied drive, this leads to the appearance of
a skyrmion Hall angle that changes sign when the field is reversed \cite{Nagaosa13,EverschorSitte18,Reichhardt22,Yang24}.
For skyrmions moving in asymmetric channels, the diode effects and rectification are the same for positive or negative fields \cite{Souza21,Souza22,Souza22a}.
Here, we propose a new type of geometry in which the transport response
of the skyrmions differs
for positive and negative magnetic fields even when the
applied current direction remains unchanged.
Since this geometry does not produce a diode effect under reversed
current, but only under reversed magnetic field, we term it
a ``magnetic diode'' effect, in analogy to the usual diode effect.
This magnetic diode effect occurs for skyrmions driven with spin-transfer torque (STT) and spin-orbit torque (SOT).

In Fig.~\ref{fig:1}(a,b), we illustrate the sample with STT driving,
while in Fig.~\ref{fig:1}(c,d) we show the sample with SOT driving.
The system consists of a sawtooth geometry where the asymmetry is opposite
on the top and bottom of the channel.
The gray regions contain high perpendicular anisotropy defects.
In this geometry, for both STT and SOT driving,
if the magnetic field is held fixed,
the velocity response of the skyrmion is the same for a fixed amplitude
current applied in the positive and negative directions,
since the asymmetry of the top and bottom sides
of the channel cancels. Therefore, no standard diode effect appears in the velocity-force (v-f) curves.
On the other hand, when the magnetic field is reversed,
there is a strong non-reciprocal response due to the Magnus force,
which breaks the symmetry by forcing the skyrmion to preferentially
interact with one side of the channel, leading to a diode-like effect
as a function of the sign of the magnetic field.
The STT and SOT produce different types of non-reciprocal velocity curves as a function of magnetic field.
This magnetic diode effect could be used as a new element for skyrmion-based devices or other devices where there is a Magnus effect.

\section{Simulation}

\begin{figure}
  \centering
  \includegraphics[width=\columnwidth]{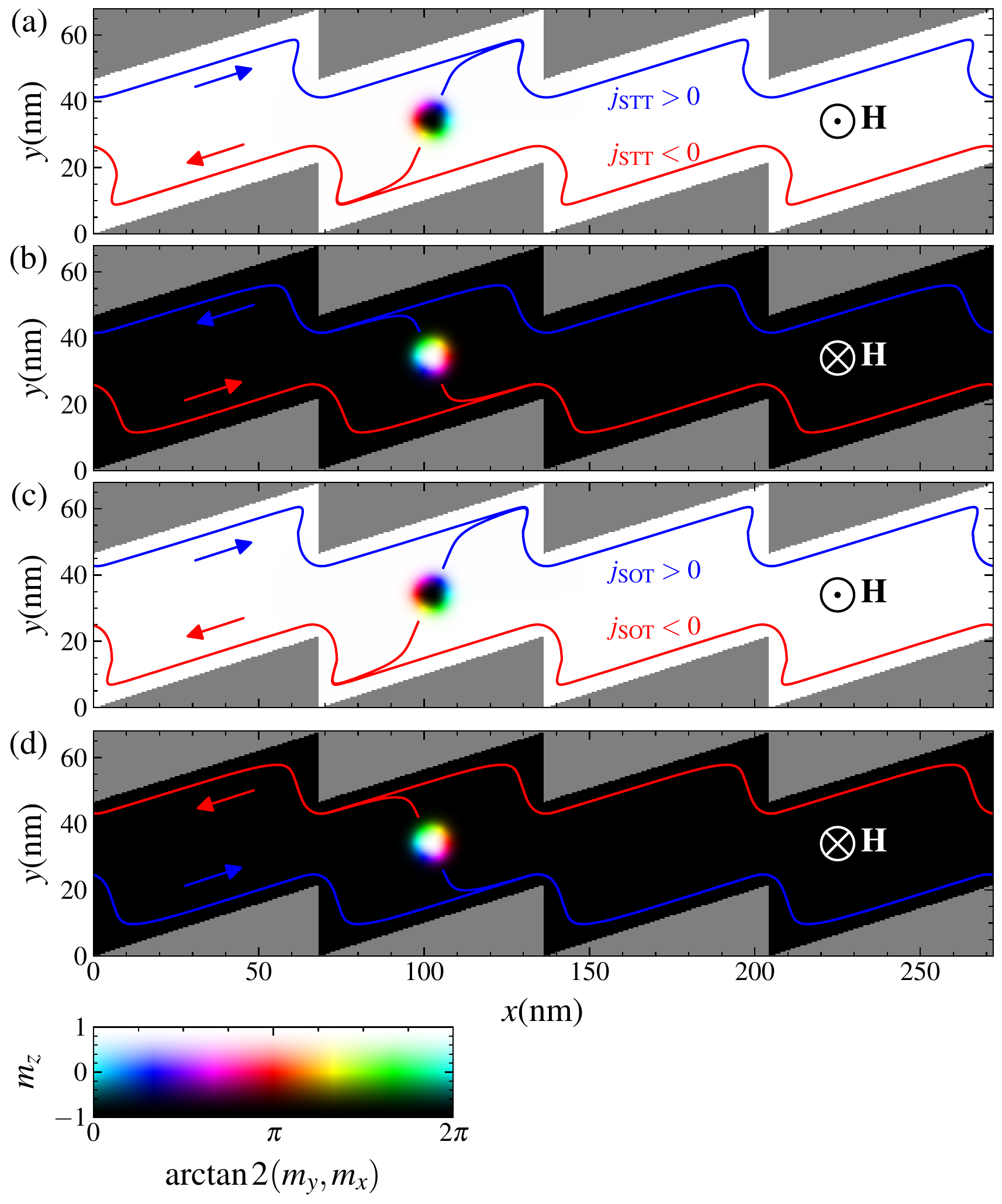}
  \caption{Illustration of the simulation system.
    Grey regions are composed of high PMA defects
    with $K_D=5J$. (a, c) A $Q=-1$ skyrmion is
    stabilized with a $+\hat{\bf z}$ magnetic field
    ${\bf H}$. (b, d) A $Q=+1$ skyrmion is stabilized with
    a $-\hat{\bf z}$ magnetic field ${\bf H}$.
    An STT current is applied in (a, b) and
    an SOT current is applied in (c,d).
    The blue (red) lines and arrows indicate the skyrmion motion
    under positive (negative) currents $j$.
    The skyrmions are colored according to their local $m_z$ values.
  }
  \label{fig:1}
\end{figure}

As shown in
Fig.~\ref{fig:1}, our system consists of a channel with a
sawtooth-like barrier, where the direction of the sawtooth is reversed on one side of the channel compared to the other.
The gray regions contain defects with high perpendicular magnetic anisotropy $K_D=5J$, such that the skyrmion will not enter the gray areas.
A $Q=\pm1$ skyrmion is placed in the channel and stabilized with a $\mp\hat{\mathbf{z}}$ magnetic field $\mathbf{H}$. We then apply a current in the positive ($j > 0$) or negative ($j < 0$) direction. We measure the skyrmion velocity as a function of $j$ for different magnetic fields.
In previous work, a sawtooth geometry was used in which the asymmetry was
the same on the top and bottom of the channel, rather than reversed as in
the present case, and it was shown that this geometry produced
a diode effect under a dc driving current and
a ratchet effect under an ac driving  current \cite{Souza21,Souza22,Souza22a},
similar to behavior that has been observed in superconducting
vortices \cite{Yu07,Wambaugh99} and
colloidal systems \cite{Reichhardt18}.

We simulate the system using atomistic simulations, which capture the dynamics of
individual atomic magnetic moments \cite{Evans18}, to
model a ferromagnetic ultrathin film capable of holding
N{\'e}el skyrmions. Our sample has dimensions of 272 nm $\times$ 68 nm with
periodic boundary conditions along the $x$ direction,
and we perform our simulations
at zero temperature $T=0$~K.

The Hamiltonian governing the atomistic dynamics is given by
\cite{Iwasaki13, Evans18, Iwasaki13a}:
\begin{eqnarray}\label{eq:1}
  \mathcal{H}=&-\sum_{\langle i,
    j\rangle}J_{ij}\mathbf{m}_i\cdot\mathbf{m}_j -\sum_{\langle i,
    j\rangle}\mathbf{D}_{ij}\cdot\left(\mathbf{m}_i\times\mathbf{m}_j\right)\\\nonumber
  &-\sum_i\mu\mathbf{H}\cdot\mathbf{m}_i - \sum_{i} K_i\left(\mathbf{m}_i\cdot\hat{\mathbf{z}}\right)^2 \ . \\\nonumber
\end{eqnarray}

The ultrathin film is modeled as a square arrangement of atoms with a
lattice constant $a=0.5$ nm. The first term on the right hand side
of Eq.~\ref{eq:1} is
the exchange interaction with an exchange constant of $J_{ij}=J$
between magnetic moments $i$ and $j$. The second term is the
interfacial Dzyaloshinskii–Moriya interaction, where
$\mathbf{D}_{ij}=D\mathbf{\hat{z}}\times\mathbf{\hat{r}}_{ij}$ is the
Dzyaloshinskii–Moriya vector between magnetic moments $i$ and $j$ and
$\mathbf{\hat{r}}_{ij}$ is the unit distance vector between sites $i$
and $j$. Here, $\langle i, j\rangle$ indicates that the sum is over
only the first neighbors of the $i$th magnetic moment. The third term
is the Zeeman interaction with an applied external magnetic field
$\mathbf{H}$. Here $\mu=g\mu_B$ is the magnitude of the magnetic
moment, $g=|g_e|=2.002$ is the electron $g$-factor, and
$\mu_B=9.27\times10^{-24}$~J~T$^{-1}$ is the Bohr magneton. The last
term represents the sample perpendicular magnetic anisotropy (PMA),
where we use two PMA constants: $K=01J$ for $i\notin P$, and $K=5J$
for $i\in P$, with $P$ being the set of defects composing the
defect arrangement shown in Fig.~\ref{fig:1}.
In ultrathin films, long-range dipolar interactions
act as a PMA (see Supplemental Material of Wang {\it et al.} \cite{Wang18}),
and therefore merely effectively shift the PMA values.

The time evolution of atomic magnetic moments is obtained using the
Landau-Lifshitz-Gilbert (LLG)
equation \cite{Seki16, Gilbert04}:
\begin{equation}\label{eq:2}
  \frac{\partial\mathbf{m}_i}{\partial
    t}=-\gamma\mathbf{m}_i\times\mathbf{H}^\text{eff}_i
  +\alpha\mathbf{m}_i\times\frac{\partial\mathbf{m}_i}{\partial t}
  +\frac{pa^3}{2e}\left(\mathbf{j}\cdot\nabla\right)\mathbf{m}_i \ .
\end{equation}
Here $\gamma=1.76\times10^{11}~$T$^{-1}$~s$^{-1}$ is the electron
gyromagnetic ratio,
$\mathbf{H}^\text{eff}_i=-\frac{1}{\mu}\frac{\partial \mathcal{H}}{\partial \mathbf{m}_i}$
is the effective magnetic field including all interactions from the Hamiltonian, $\alpha$ is the
phenomenological damping introduced by Gilbert, and the last term is
the adiabatic spin-transfer-torque (STT) caused by application of an in
plane spin polarized current, where $p$ is the spin polarization, $e$
the electron charge, and $\mathbf{j}=j\hat{\mathbf{y}}$ the applied
current density. Use of this STT expression implies that the
conduction electron spins are always parallel to the magnetic moments
$\mathbf{m}$ \cite{Iwasaki13,Zang11}.
In this work the adiabatic STT contribution
corresponds to dragging forces perpendicular to ${\bf j}$ \cite{Iwasaki13a, Feilhauer20},
so dragging forces are along $\hat{\bf x}$ in our system.
It is possible to include non-adiabatic STT contributions in Eq.~\ref{eq:2};
however, non-adiabatic STT does not appreciably affect
the dynamics of nanoscale skyrmions at small driving forces
\cite{Litzius17}, which is the regime we consider.
The other external torque contribution is given by
$\boldsymbol{\tau}_\text{SOT}=\hbar\gamma P a^2/(2e\mu){\bf m}_i\times\left(\hat{{\bf z}}\times{{\bf j}}_\text{SOT}\right)\times{\bf m}_i$,
which is caused by a spin-orbit-torque type of applied current;
in contrast to STTs, the SOT dragging force is along ${\bf j}_\text{SOT}$.
Here ${\bf j}_\text{SOT}=j\hat{\bf x}$ is the applied current density, $P=1$ is the current polarization,
and $e$ is the electron charge.

The skyrmion velocity is computed using the emergent electromagnetic
fields \cite{Seki16, Schulz12}:
\begin{eqnarray}\label{eq:5}
  E^\text{em}_i=\frac{\hbar}{e}\mathbf{m}\cdot\left(\frac{\partial
    \mathbf{m}}{\partial i}\times\frac{\partial \mathbf{m}}{\partial
    t}\right)\\ B^\text{em}_i=\frac{\hbar}{2e}\varepsilon_{ijk}\mathbf{m}\cdot\left(\frac{\partial
    \mathbf{m}}{\partial j}\times\frac{\partial \mathbf{m}}{\partial
    k}\right) \ ,
\end{eqnarray}
where $\varepsilon_{ijk}$ is the totally anti-symmetric tensor. The
skyrmion drift velocity, $\mathbf{v}_d$, is then calculated using
$\mathbf{E}^\text{em}=-\mathbf{v}_d\times\mathbf{B}^\text{em}$.

We fix the following values in our simulations:
$J=1$ meV and $D=0.2J$. The values $\mu H=0.5D^2/J$ and $\alpha=0.3$
are used unless otherwise mentioned.
The numerical integration of Eq.~\ref{eq:2} is
performed using a fourth order Runge-Kutta method.
To ensure a steady state for measurement we evolve
Eq.~\ref{eq:2} over 200 ns.

\section{Results}

\begin{figure}
  \centering
  \includegraphics[width=\columnwidth]{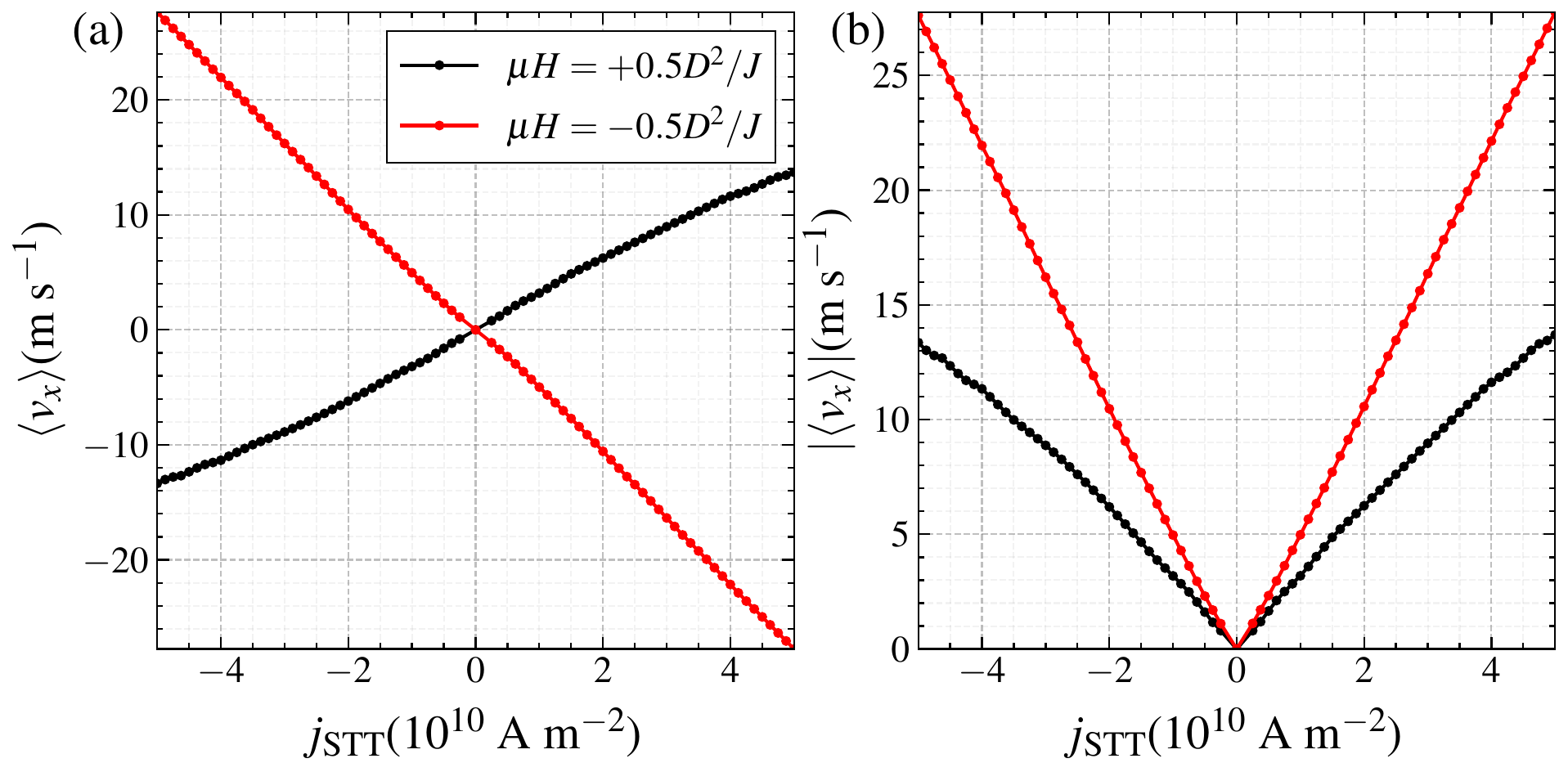}
  \caption{(a) Average skyrmion velocity
$\langle v_x\rangle$ vs
applied current $j_\text{STT}$ with STT driving under positive, $\mu H=+0.5D^2/J$  (black),
	and negative, $\mu H=-0.5D^2/J$ (red), magnetic fields,
    for a system with $\alpha=0.3$.
(b) The corresponding absolute average skyrmion velocity
    $|\langle v_x\rangle|$ vs
    applied current $j_\text{STT}$ under positive, $\mu H=+0.5D^2/J$ (black),
    and negative, $\mu H=-0.5D^2/J$ (red), magnetic fields.
  }
  \label{fig:2}
\end{figure}

\begin{figure}
  \centering
  \includegraphics[width=\columnwidth]{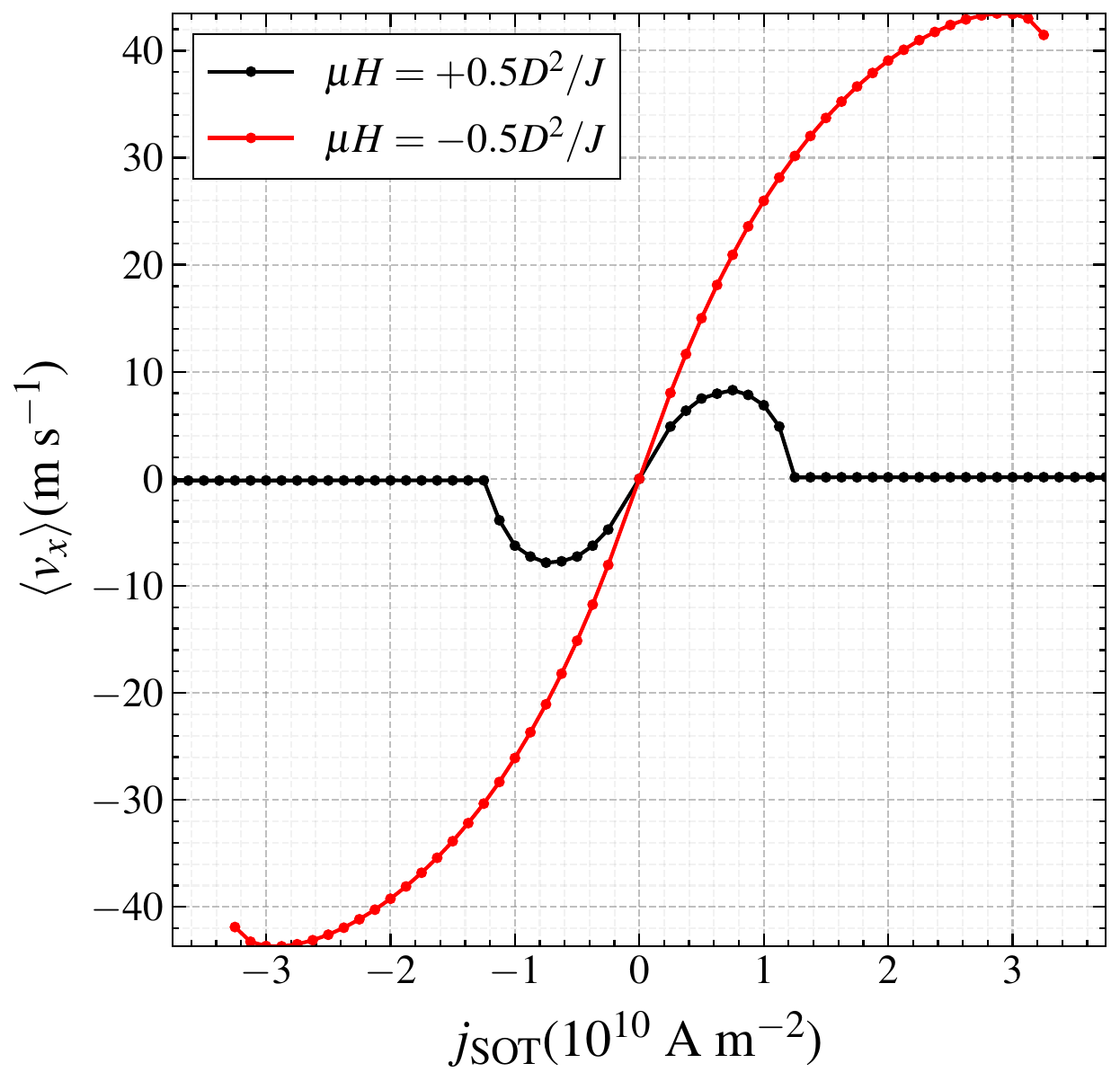}
  \caption{Average skyrmion velocity $\langle v_x\rangle$ vs
applied current $j_\text{SOT}$ with SOT driving under positive, $\mu H=+0.5D^2/J$  (black),
    and negative, $\mu H=-0.5D^2/J$ (red), magnetic fields,
    for a system with $\alpha=0.3$.
  }
  \label{fig:3}
\end{figure}

In Fig.~\ref{fig:1}(a,b), a single skyrmion is subjected to STT driving
in our asymmetric sawtooth channel.
Since the net asymmetry of the top and bottom halves of the channel
is zero,
there is no diode or ratchet effect in this system when the
magnetic field is held fixed.
Due to the Magnus force, however, the response can change when
the magnetic field is reversed.
In Fig.~\ref{fig:1}(a), with the magnetic field in the positive $z$-direction
and $j_\text{STT} > 0$, the Magnus force causes the skyrmion
to traverse the upper part of the channel in the $+x$-direction,
meaning that the skyrmion is moving along the hard asymmetry direction of
this side of the channel.
For $j_\text{STT} < 0$, the skyrmion
moves along the bottom part of the channel
in the $-x$-direction and is again traveling in the hard asymmetry
direction,
so there is no diode effect when the current is reversed but the magnetic
field is unchanged.
When the field is reversed into the negative $z$-direction,
as shown in Fig.~\ref{fig:1}(b),
for $j_\text{STT} > 0$,
the skyrmion now moves in the $-x$ direction along the upper part
of the channel, which is the easy asymmetry direction,
and for $j_\text{STT} < 0$, the skyrmion moves in the positive $x$-direction
along the bottom part of the channel, again
in the easy asymmetry direction.
The skyrmion velocity changes sign but does not change in magnitude when
the current is reversed; however, the magnitude of the velocity is larger
for negative field than for positive field, meaning that the flow resistance
depends on the magnetic field direction.

In Fig.~\ref{fig:1}(c), we show the skyrmion motion under SOT driving with the magnetic field applied in the positive $z$-direction.
When $j_\text{SOT} > 0$, the skyrmion moves in the $+x$-direction or hard
asymmetry direction of the upper part of the channel, and similarly
when $j_\text{SOT} < 0$, the skyrmion moves in the $-x$-direction or hard
asymmetry direction of the bottom part of the channel.
We again find that there is no diode effect
produced by changing the direction of the current.
When the field is in the $-z$-direction, as shown in Fig.~\ref{fig:1}(d),
for $j_\text{SOT} > 0$, the skyrmion moves in the positive $+x$-direction
but is now on the bottom part of the channel and moving in the
easy asymmetry direction.
For $j_\text{SOT} < 0$, the skyrmion moves in the $-x$-direction or
easy asymmetry direction of the upper part of the channel.
In the SOT case, if the current direction is held fixed,
the skyrmion moves in the same direction for positive or negative
magnetic fields,
but for the positive field,
the skyrmion moves along the hard direction of the substrate asymmetry,
and therefore travels more slowly than under negative magnetic field,
when it travels in the easy direction of the substrate asymmetry.

To quantify the non-reciprocal effects as a function of field,
in Fig.~\ref{fig:2}(a), we plot $\langle v_x \rangle$ versus
$j_{\text{STT}}$ for the system in Fig.~\ref{fig:1}(a,b)
with $\alpha = 0.3$
at $\mu H = +0.5D^2/J$ and $\mu H = -0.5D^2/J$.
For fixed field, there is no diode effect in the velocity-force curves,
since each is symmetric about zero.
However, for the different fields, two effects are clear.
The direction of the velocity changes sign from positive for a positive field to negative for a negative field,
and the absolute value of the velocity at a given value of $j$ is
greater for the negative field,
since the skyrmion can move along an easy substrate asymmetry
direction, as seen in Fig.~\ref{fig:1}(a,b).
In Fig.~\ref{fig:2}(b), we plot the absolute average skyrmion velocity
$|\langle v_x \rangle|$ versus the applied current $j_\text{STT}$
under positive, $\mu H = +0.5D^2/J$, and negative, $\mu H = -0.5D^2/J$,
magnetic fields.
Here, it can be more clearly seen that the skyrmions
are moving more rapidly for $\mu H = -0.5D^2/J$ than for $\mu H = +0.5D^2/J$.
For the STT driving, there is no threshold current for skyrmion motion for either positive or negative drives, as the skyrmion is able to overcome
the sawtooth potential barrier
in the hard direction at any finite drive.

\begin{figure}
  \centering
  \includegraphics[width=\columnwidth]{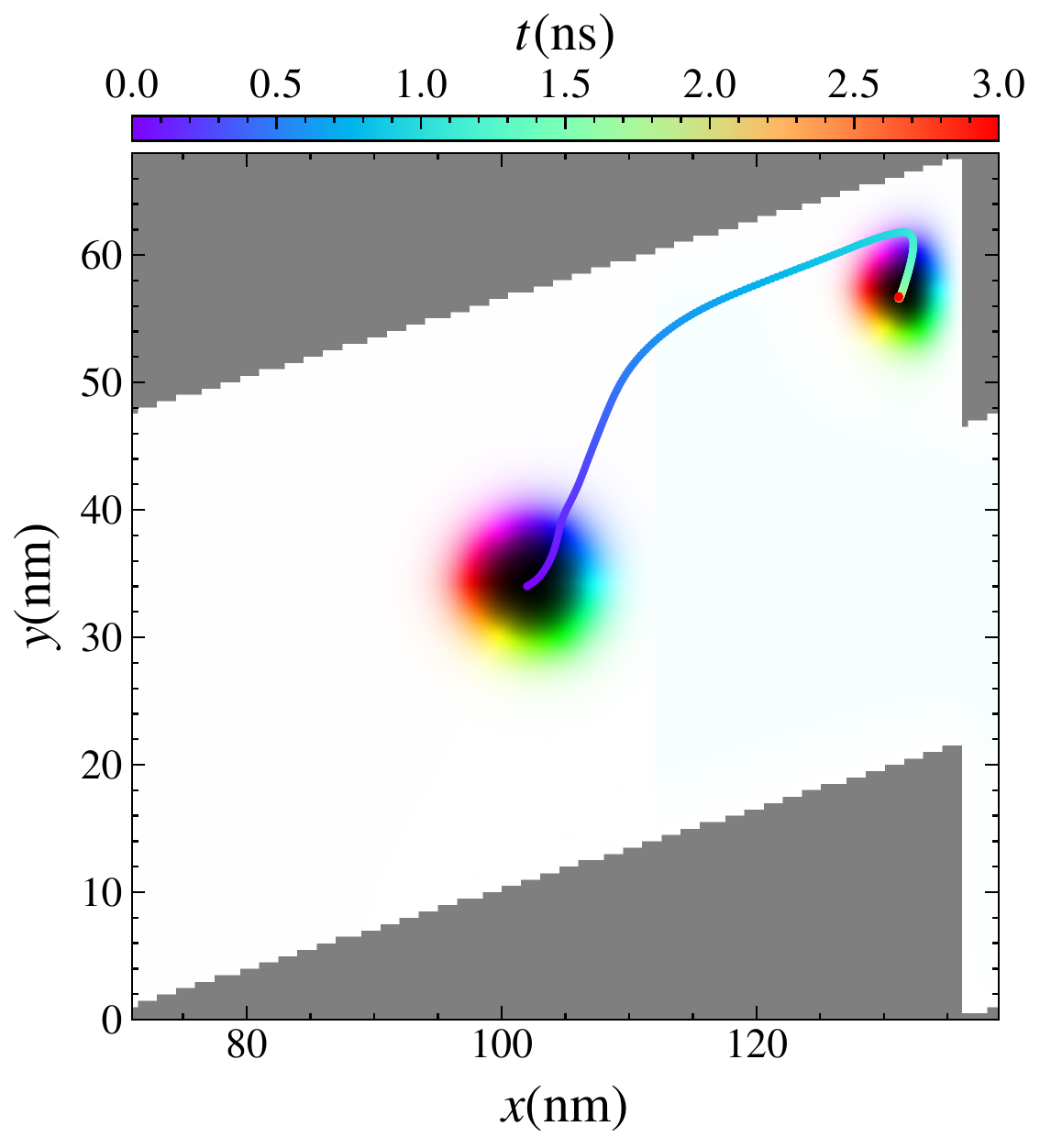}
  \caption{Image of skyrmion motion from the initial position at the
    center of the channel to the final trapped position in the corner of the
    well for the system from Fig.~\ref{fig:3} with SOT driving
    at
    $j_\text{SOT}=2\times 10^{10}$ A m$^{-2}$
    and positive magnetic field
    $\mu H = +0.5 D^2/J$.
    The colorbar indicates the passage of time but is cut off
    at $t=3$~ns for visualization purposes; once trapped,
    the skyrmion remains pinned for the duration of the simulation
    up to $t=200$~ns.
  }
  \label{fig:4}
\end{figure}

In Fig.~\ref{fig:3}, we plot the skyrmion velocity versus
current curves for SOT driving at $\mu H = +0.5D^2/J$ and
$\mu H = -0.5D^2/J$.
In this case, the velocity for $j > 0$ is always positive
for both positive and negative fields, and at low $j$, the
response is similar for both field directions. For
intermediate drive amplitudes, however,
the velocity is much larger for the negative magnetic field
than for the positive magnetic field,
and for $j > 1.2\times10^{10}$~A~m$^{-2}$, the skyrmion
undergoes a reentrant
pinning effect for the positive field when its size is partially reduced
and it
becomes trapped in the corner of the channel,
as shown in Fig.~\ref{fig:4}.
There is a small reduction in the velocity at
higher currents for the negative field,
but there is a wide range of $j$ over which
the velocity is finite for negative magnetic field
but zero for positive magnetic field.
Figure~\ref{fig:3} also shows that for a fixed field direction,
no diode effect can be produced solely by changing the direction
of the current.

\begin{figure}
  \centering
  \includegraphics[width=\columnwidth]{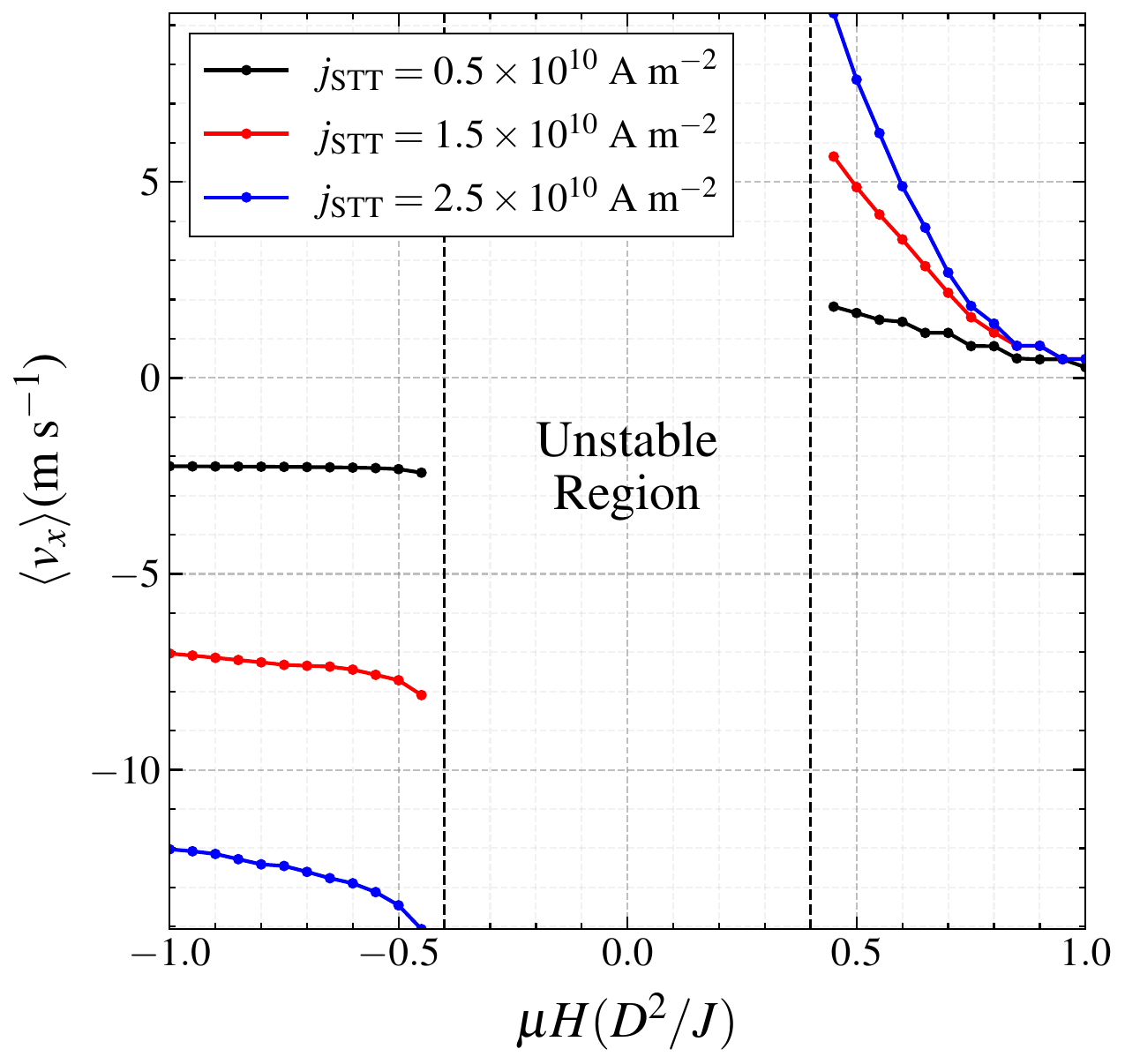}
  \caption{Average skyrmion velocity
    $\langle v_x\rangle$ vs applied magnetic
    field $\mu H$ for a system with $\alpha=0.3$ and STT driving at
    $j_\text{STT}=+2.5\times10^{10}$~A~m$^{-2}$ (blue),
    $j_\text{STT}=+1.5\times10^{10}$~A~m$^{-2}$ (red), and
    $j_\text{STT}=+0.5\times10^{10}$~A~m$^{-2}$ (black).
  }
  \label{fig:5}
\end{figure}

To make the analogy to a diode more clear, in Fig.~\ref{fig:5} we plot
$\langle v_x \rangle$ versus the applied magnetic field $\mu H$
for the STT system from Fig.~\ref{fig:2}
at fixed current densities of $j = 2.5 \times 10^{10}$~A~m$^{-2}$,
$j = 1.5 \times 10^{10}$~A~m$^{-2}$, and $j = 0.5 \times 10^{10}$~A~m$^{-2}$.
At low fields, $|\mu H| < 0.4 D^2/J$, the skyrmion is not stable. For $\mu H > 0.4D^2/J$ at $j = 2.5 \times 10^{10}$~A~m$^{-2}$,
the velocity drops with increasing field
from $\langle v_x \rangle = 10$~m~s$^{-1}$ to nearly
zero at $\mu H = 1.0 D^2/J$,
while for $\mu H < -0.4 D^2/J$,
the magnitude of the velocity
decreases from
$\langle v_x \rangle = -14$~m~s$^{-1}$
to $\langle v_x \rangle = -12$~m~s$^{-1}$ near $\mu H = -1.0 D^2/J$.
As a result, the absolute value of the skyrmion velocity
is much higher for the negative fields than for positive fields
of the same magnitude.
Additionally, at $\mu H = \pm 1.0 D^2/J$,
the system exhibits a near-perfect magnetic diode effect,
where the skyrmion has a finite velocity for negative fields
but is pinned or has zero velocity for positive magnetic fields.
For $j = 1.5 \times 10^{10}$~A~m$^{-2}$, the diode effect is reduced,
since the velocity for positive fields ranges from a maximum of
6~m~s$^{-1}$ before dropping close to zero at $\mu H = 1.0$,
while for negative fields,
the velocity magnitude drops from
$\langle v_x \rangle = -8$~m~s$^{-1}$ to $-7$~m~s$^{-1}$ at $\mu H = -1.0 D^2/J$. For $j = 0.5 \times 10^{10}$~A~m$^{-2}$,
the diode efficiency is even more strongly reduced.
The results in Fig.~\ref{fig:5} indicate that the diode efficiency can be
increased by applying
larger fields and higher $j_\text{STT}$.

\begin{figure}
  \centering
  \includegraphics[width=\columnwidth]{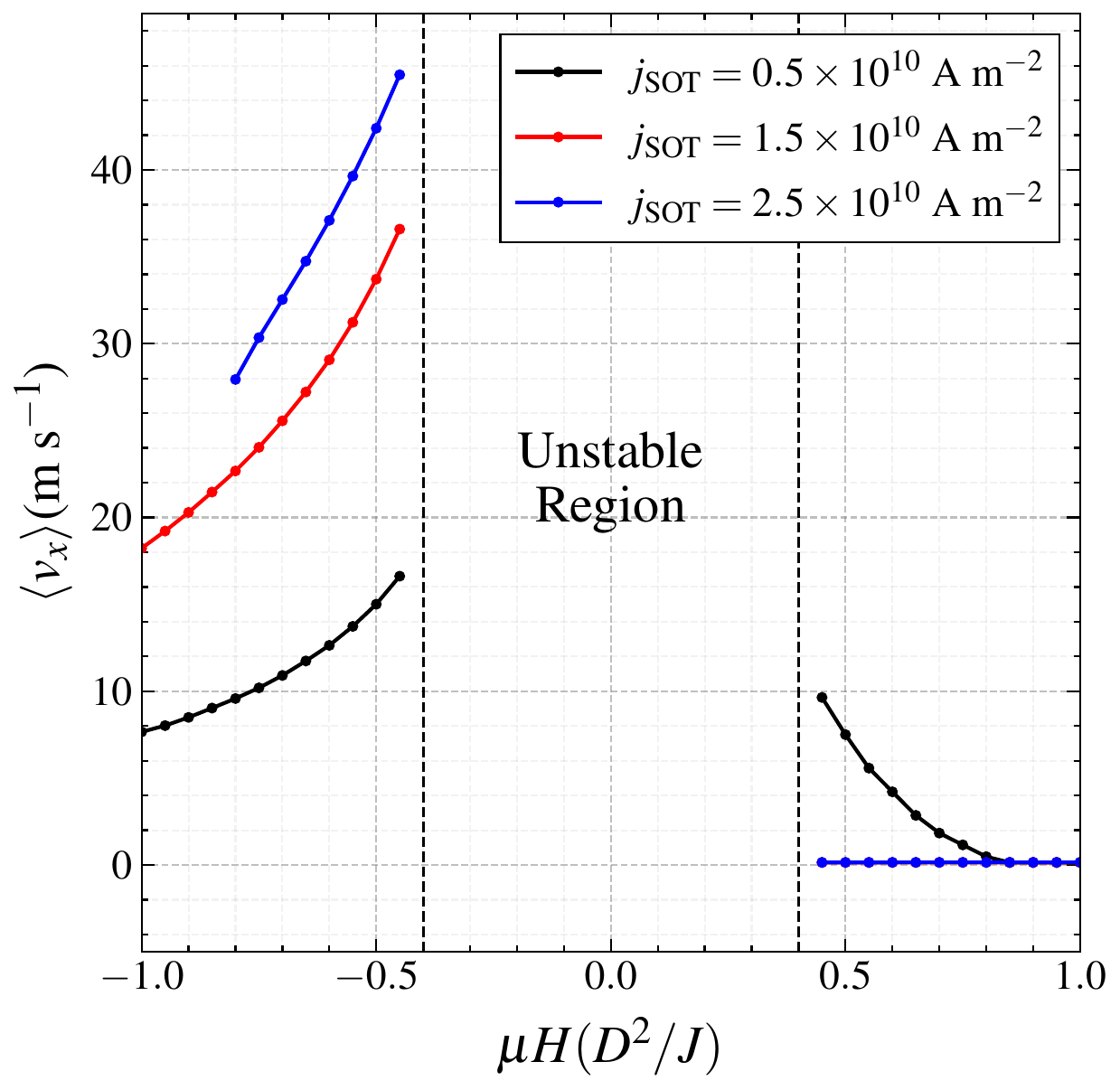}
  \caption{Average skyrmion velocity
    $\langle v_x \rangle$ vs applied magnetic field $\mu H$
    for a system with $\alpha = 0.3$ and SOT driving at
    $j_\text{SOT} = +2.5 \times 10^{10}$~A~m$^{-2}$ (blue),
    $j_\text{SOT} = +1.5 \times 10^{10}$~A~m$^{-2}$ (red),
    and $j_\text{SOT} = +0.5 \times 10^{10}$~A~m$^{-2}$ (black).
  }
  \label{fig:6}
\end{figure}

Figure~\ref{fig:6} shows $\langle v_x \rangle$ versus $\mu H$
for SOT driving at $j = +2.5 \times 10^{10}$~A~m$^2$,
$j = +1.5 \times 10^{10}$~A~m$^2$, and $j = +0.5 \times 10^{10}$~A~m$^2$.
At $j = +2.5 \times 10^{10}$~A~m$^2$,
the skyrmion becomes pinned for positive magnetic fields
and $\langle v_x \rangle$ = 0~m~s$^{-1}$,
while for negative fields,
the velocity drops with increasing field magnitude
from $\langle v_x \rangle = 45$~m~s$^{-1}$
to $28$~m~s$^{-1}$, indicating a perfect diode effect.
For $j = +1.5 \times 10^{10}$~A~m$^2$,
$\langle v_x \rangle = 0$~m~s$^{-1}$ for positive field,
and the velocity is finite but reduced for negative field.
For $j = +0.5 \times 10^{10}$~A~m$^2$,
the velocity is finite for both positive and negative fields,
but remains asymmetric so there is still a diode effect.
In general, for both STT and SOT driving,
the magnetic diode efficiency is higher for larger fields and currents.

\begin{figure}
  \centering
  \includegraphics[width=\columnwidth]{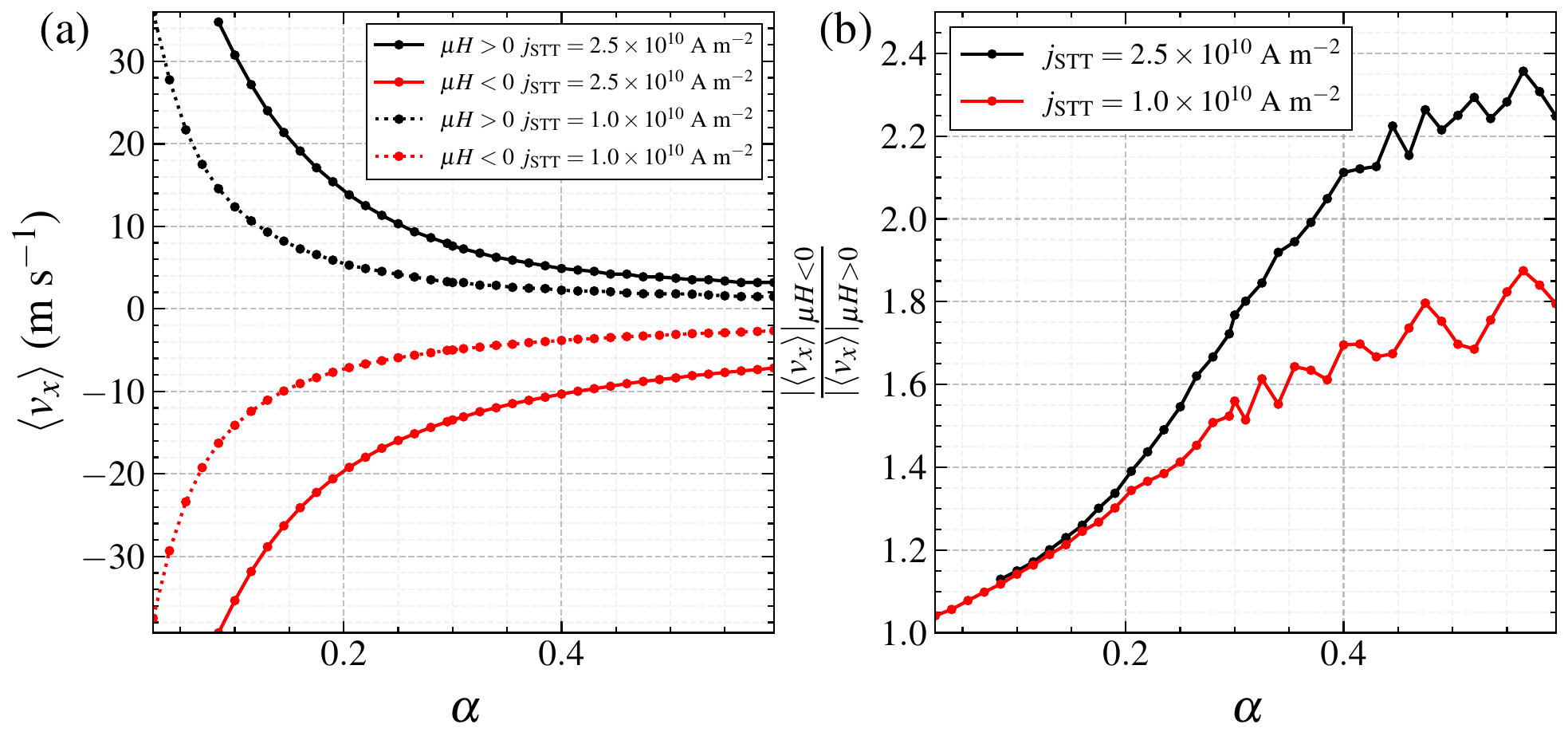}
  \caption{
(a) Average skyrmion velocity $\langle v_x\rangle$ versus the Gilbert damping $\alpha$ for $|\mu H| = 0.5D^2/J$ and different current densities
for STT driving: $j_{\text{STT}} = +2.5 \times 10^{10}$~A~m$^{-2}$ (black solid) and $j_{\text{STT}} = +1.0 \times 10^{10}$~A~m$^{-2}$ (black dashed) with $\mu H=+0.5D^2/J$. The red solid line is for $j_{\text{STT}} = +2.5 \times 10^{10}$~A~m$^{-2}$, and the red dashed line is for $j_{\text{STT}} = +1.0 \times 10^{10}$~A~m$^{-2}$ with $\mu H = -0.5D^2/J$.
    (b) Ratio of the absolute value of the average skyrmion velocity along the $x$-axis for negative and positive fields, $\frac{|\langle v_x\rangle|_{\mu H<0}}{|\langle v_x\rangle|_{\mu H>0}}$, versus the Gilbert damping $\alpha$ for systems with different current densities and $|\mu H| = 0.5D^2/J$.
    We plot only positive $j$ systems, but the negative $j$ behavior is very similar to the positive $j$ behavior.
  }
  \label{fig:7}
\end{figure}

In Fig.~\ref{fig:7}(a), we plot $\langle v_x\rangle$ versus the Gilbert damping $\alpha$ for a system with $j_{\text{STT}} = +2.5 \times 10^{10}$~A~m$^{-2}$ and $j_{\text{STT}} = +1.0 \times 10^{10}$~A~m$^{-2}$ for $\mu H = 0.5D^2/J$,
along with $j_{\text{STT}} = +2.5 \times 10^{10}$~A~m$^{-2}$ and $j_{\text{STT}} = +1.0 \times 10^{10}$~A~m$^{-2}$ for $\mu H = -0.5D^2/J$.
With increasing $\alpha$, the
magnitude of the velocity for both directions of the field decreases.
To better characterize the diode efficiency,
in Fig.~\ref{fig:7}(b), we plot
the ratio $\frac{|\langle v_x\rangle|_{\mu H<0}}{|\langle v_x\rangle|_{\mu H>0}}$ versus $\alpha$ for $j_{\text{STT}} = +2.5 \times 10^{10}$~A~m$^{-2}$
and $j_{\text{STT}} = +1.0 \times 10^{10}$~A~m$^{-2}$.
For $j_{\text{STT}} = +2.5 \times 10^{10}$~A~m$^{-2}$, the diode efficiency starts off near 1.0 for small $\alpha$, indicating a lack of a diode effect, and then increases, reaching a maximum of nearly 2.2 for large $\alpha$.
This indicates that the diode effect should be robust for relativistic values of the damping and can even be enhanced at larger $\alpha$.
For $j_{\text{STT}} = +1.0 \times 10^{10}$~A~m$^{-2}$, a similar trend
appears; however, the maximum diode efficacy is reduced to 1.9.
We plot only positive $j$ systems, but the negative $j$ behavior is very similar to the positive $j$ behavior.

\begin{figure}
  \centering
  \includegraphics[width=\columnwidth]{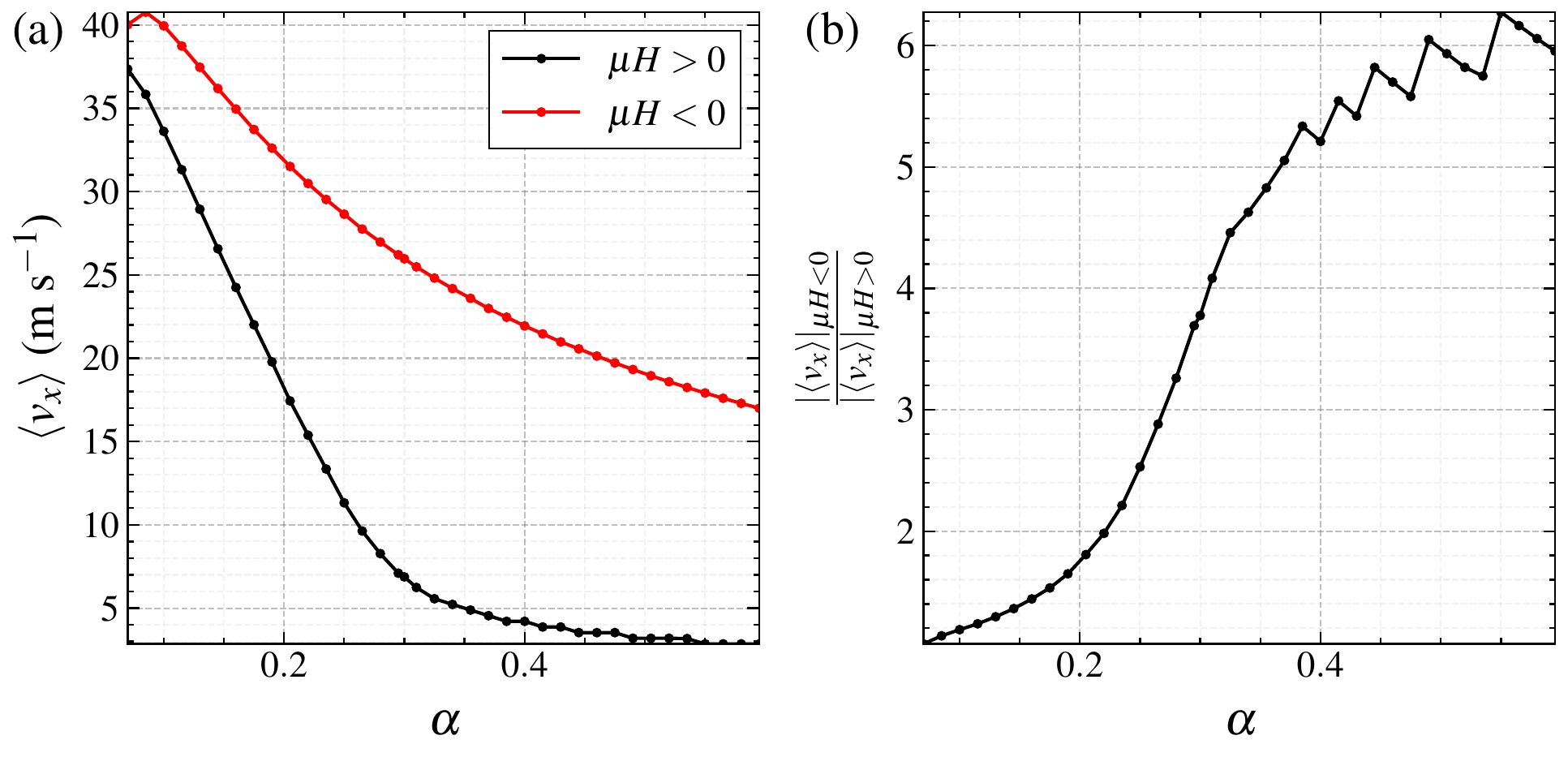}
  \caption{
    (a) Average skyrmion velocity $\langle v_x \rangle$ vs the Gilbert damping $\alpha$ for $|\mu H| = 0.5 D^2/J$ with SOT driving with $j_{\text{SOT}} = 1.0 \times 10^{10}$~A~m$^{-2}$ (black) and $j_{\text{SOT}} = +1.0 \times 10^{10}$~A~m$^{-2}$ (red) for $\mu H = -0.5 D^2/J$.
    (b) Ratio of the absolute value of the average skyrmion velocity for negative and positive magnetic fields, $\frac{|\langle v_x \rangle|_{\mu H < 0}}{|\langle v_x \rangle|_{\mu H > 0}}$, vs the Gilbert damping $\alpha$.
  }
  \label{fig:8}
\end{figure}

In Fig.~\ref{fig:8}(a), we plot $\langle v_x \rangle$ versus the Gilbert damping $\alpha$ for a system with $j_{\text{SOT}} = +1.0 \times 10^{10}$~A~m$^{-2}$ at $\mu H = +0.5 D^2/J$ and $\mu H = -0.5 D^2/J$. In this case, the velocity is positive for both positive and negative fields, and drops with increasing $\alpha$; however, it falls off more rapidly for the positive fields.
Figure~\ref{fig:8}(b) shows the corresponding $\frac{|\langle v_x \rangle|{\mu H < 0}}{|\langle v_x \rangle|{\mu H > 0}}$ versus $\alpha$, which starts near 1.0 (no diode effect) at small $\alpha$ and increases with increasing $\alpha$.
There rate of increase slows
for $\alpha > 0.4$, and the ratio asymptotes to a value
near 6 at large $\alpha$.
This shows that the diode effect is stronger for the SOT driving
than for the STT driving.
We find similar effects for other values of $\mu H$ and $j$,
indicating
that the magnetic field effect in this geometry is robust for a range of parameters.

\section{Discussion}
Our results focused on skyrmions where there is a Magnus effect,
but we expect that a similar magnetic diode effect
could arise in other systems with Magnus-like dynamics in a similar
geometry. Such systems could include Wigner crystals in magnetic fields or certain chiral active matter
systems \cite{Ai16}.
Some future directions would be to
consider collective effects for multiple interacting skyrmions,
as well as the effects of other geometries on
two-dimensional arrays with modified anisotropy.
In this work, we considered atomistic models, but it would also be
interesting to see if similar effects can be captured using
Thiele or Langevin-based approaches \cite{Lin13}.

\section{Conclusion}

Using atomistic simulations, we have shown that a skyrmion moving in a channel with two sawtooth substrates that have reversed asymmetries can exhibit what we call a magnetic diode effect.
In a standard diode effect,
the velocity of the skyrmion is be non-reciprocal for opposite
driving directions.
In our geometry, for a fixed magnetic field,
there is no diode effect as a function of drive or current;
however, when the magnetic field is reversed at fixed drive,
we obtain a non-reciprocal response in the skyrmion velocity.
For systems with spin-transfer torque driving,
the velocity is reversed in direction, and the absolute value of the
velocity changes when the magnetic field is reversed.
For spin-orbit torque driving, the velocity is
in the same direction for both positive and negative
magnetic fields,
but the skyrmion is pinned for positive magnetic fields
moving for negative magnetic fields,
producing a perfect magnetic diode effect.
We show that the magnetic diode
effect is robust for a wide range of magnetic fields,
currents, and values of the Gilbert damping.
The magnetic diode effect
arises due to the Magnus force on the skyrmions,
which causes the skyrmion to interact more strongly
with one side of the channel than the other under a magnetic field.
For one magnetic field polarity, the skyrmion travels along
a hard direction of the substrate asymmetry,
while for the opposite field
polarity, it travels along an easy direction of the substrate asymmetry.
This magnetic field diode could be used
to create a new type of
logic element for applications using skyrmions.
Additionally, this type of magnetic diode
could be relevant for other systems with a strong Magnus force.

\begin{acknowledgments}
  This work was supported by the US Department of Energy through the Los Alamos
  National Laboratory. Los Alamos National Laboratory is operated by
  Triad National Security, LLC, for the National Nuclear Security
  Administration of the U. S. Department of Energy (Contract
  No. 892333218NCA000001).
  %
  %
  We would like to thank Dr. Pablo Antonio Venegas for providing the computational
  resources used in this work through Funda\c{c}\~ao de Amparo \`a Pesquisa
  do Estado de S\~ao Paulo - FAPESP, grant 2024/02941-1.
\end{acknowledgments}

\bibliography{mybib}

\end{document}